\documentclass{webofc}
\usepackage[varg]{txfonts}   % Web of Conferences font
\usepackage{hyperref}
\usepackage{url}
\hypersetup{colorlinks=true,citecolor=blue,urlcolor=blue,linkcolor=blue}
\usepackage{graphicx}
\usepackage{subcaption}

\newcommand{\sqrts}{\sqrt{s_\mathrm{NN}}}

\begin{document}
\title{Simulating collectivity in dense baryon matter with multiple fluids}

\author{\firstname{Iurii} \lastname{Karpenko}\inst{1}\fnsep\thanks{\email{yu.karpenko@gmail.com}} \and
        \firstname{Jakub} \lastname{Cimerman} \and
        \firstname{Pasi} \lastname{Huovinen}\inst{2} \and
        \firstname{Boris} \lastname{Tom\'a\v{s}ik}\inst{1,3}
}

\institute{Faculty of Nuclear Sciences and Physical Engineering, Czech Technical University in Prague,\\ Břehová 7, Prague, Czech Republic
\and
Incubator of Scientific Excellence---Centre for Simulations of Superdense Fluids, University of Wroc\l{}aw,  Poland
\and
Univerzita Mateja Bela, Bansk\'a Bystrica, Slovakia
          }

\abstract{We report on construction of a modern multi-fluid approach to heavy-ion collisions at FAIR/BES energies (MUFFIN) and show the reproduction of basic experimental observables in Au-Au collisions in the RHIC Beam Energy Scan program. We also show the $p_T$-differential and $p_T$-integrated polarization of (anti-)$\Lambda$ hyperons. In MUFFIN simulations, we observe a strong splitting between polarizations of $\Lambda$ and anti-$\Lambda$. The splitting is driven purely by a finite baryon chemical potential.
}
\maketitle
\vspace*{-10pt}
\section{Introduction}
\label{sec-intro}
One of the key objectives of heavy-ion collision experiments at center-of-mass energies ranging from a few to a hundred GeV is to explore the characteristics of the dense baryonic matter that is formed, particularly its equation of state (EoS) and transport properties. The fluid dynamic approach plays a crucial role in this endeavor, as it allows for the flexible incorporation of different equations of state.

This approach has proven highly effective in describing nucleus-nucleus collisions at high energies, specifically for $\sqrts=200$~GeV and above. At these energy scales, the evolution is typically divided into two stages: the initial state, where hard scatterings occur and lead to the isotropization or effective fluidization of the system, followed by the fluid phase, which is governed by fluid dynamical equations.

At lower energies, however, modeling heavy-ion collisions presents a distinct challenge. The Lorentz contraction of the colliding nuclei is weaker, requiring up to several fm/c for the nuclei to fully pass through one another and for all primary nucleon-nucleon (NN) scatterings to take place. In the meantime, a dense medium may already form in regions where the first NN scatterings have occurred, even while some nucleons are still approaching their first interactions.

Multi-fluid dynamics offers an elegant, albeit phenomenological, method to account for the complex space-time evolution of nucleus-nucleus collisions at intermediate energies. In this approach, the incoming nuclei are approximated as two blobs of cold, baryon-rich fluids. The collision is then modeled as an interpenetration of these fluids, which are decelerated by frictional forces. The energy and momentum lost due to friction is used to generate a third fluid, representing the particles produced during the reaction.

In the following, we present a small selection of the basic observables, as well as polarization observables, computed in the MUFFIN (MUlti Fluid simulation for Fast IoN collisions) model. For a detailed description of the model and a broader set of key observables — such as rapidity distributions and transverse momentum spectra of various hadron types — we direct the reader to \cite{Cimerman:2023hjw}.

\vspace*{-10pt}
\section{Multi-fluid model}
\label{sec-model}
Dynamics of heavy-ion reactions in the multi-fluid model (3-fluid model specifically) is desrcibed by a set of coupled local energy-momentum and baryon charge (non-)conservation equations:
\begin{align}
    \partial_\mu T^{\mu\nu}_\mathrm{p}(x)&=-F_\mathrm{p}^\nu(x)+F_\mathrm{fp}^\nu(x), \nonumber \\
    \partial_\mu T^{\mu\nu}_\mathrm{t}(x)&=-F_\mathrm{t}^\nu(x)+F_\mathrm{ft}^\nu(x), \label{3fh-equations}\\
    \partial_\mu T^{\mu\nu}_\mathrm{f}(x)&=F_\mathrm{p}^\nu(x)+F_\mathrm{t}^\nu(x)-F_\mathrm{fp}^\nu(x)-F_\mathrm{ft}^\nu(x), \nonumber \\
    \partial_\mu N^{\mu}_\mathrm{p,t,f}(x)&=0, \nonumber
\end{align}
each equation represents evolution of a fluid with a source term describing local energy-momentum exchange with the two other fluids. The indices ``p'' and ``t'' stand for projectile and target fluids which represent the original nuclei, and ``f'' stands for the fireball fluid, which represents meson-dominated medium created in the collision. Projectile and target fluids are initialised as moving blobs with local rest frame energy $\varepsilon_0$ and baryon $n_0$ densities equal to those of ordinary nuclear matter, whereas the fireball fluid is empty at the beginning.

The first three equations (\ref{3fh-equations}) summed together correspond to exact energy-momentum conservation of the entire system:
\begin{align*}
    \partial_\mu \left[T^{\mu\nu}_{p}(x) + T^{\mu\nu}_{t}(x) + T^{\mu\nu}_{f}(x) \right]=0.
\end{align*}
The functional form of the friction terms is derived from the energy and momentum transport in NN and N$\pi$ collisions using kinetic theory. The friction terms are then scaled to take into account for shortcomings of the derivation e.g.~emergence of the QGP phase. The scaling also allows for a better fit to the experimental data.

\section{Results and discussion}\label{sec-results}
\begin{figure*}[h]
\begin{subfigure}{5.5cm}
 \includegraphics[width=\textwidth]{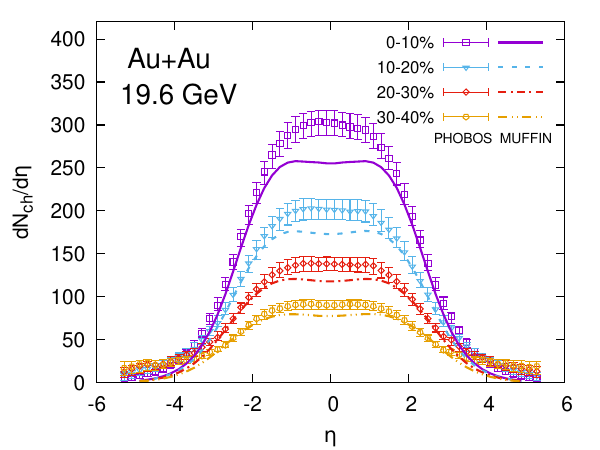}
 \includegraphics[width=\textwidth]{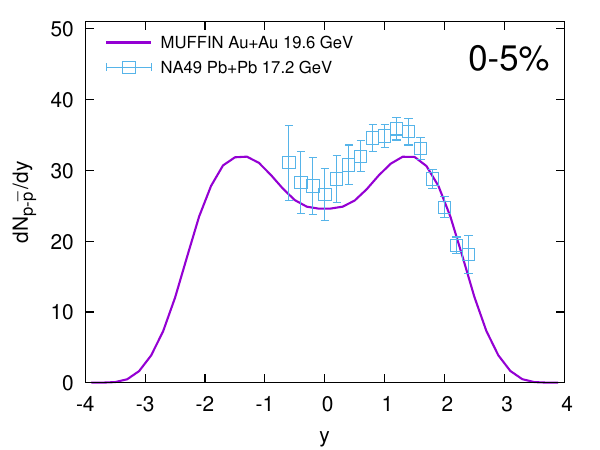}
\end{subfigure}
\begin{subfigure}{8cm}
 \includegraphics[width=\textwidth]{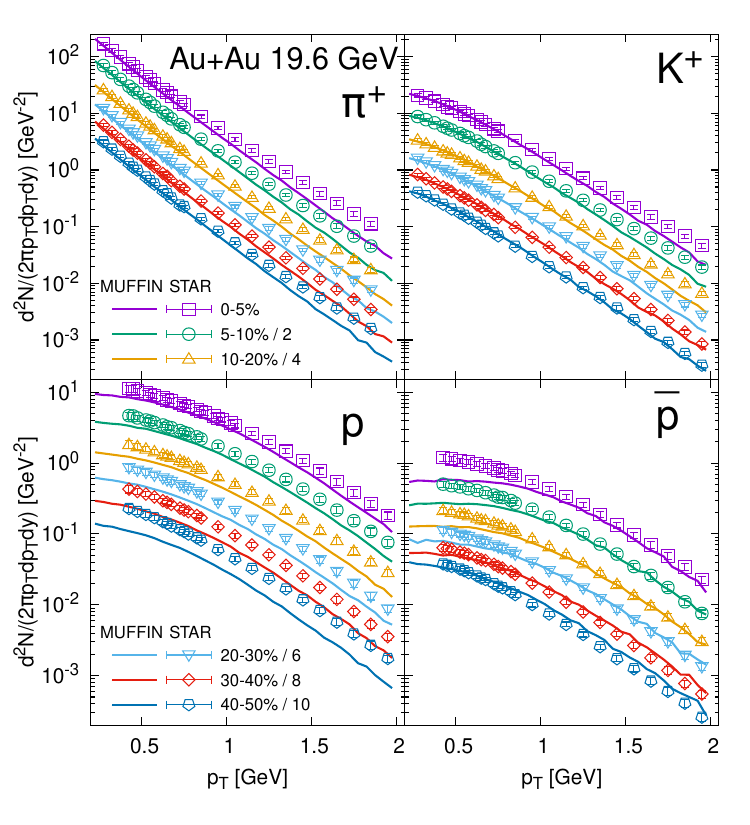}
\end{subfigure}
\caption{A selection of experimental observables from MUFFIN for Au-Au collisions at $\sqrts=7.7$~GeV: pseudorapidity distributions of all charged hadrons compared to PHOBOS data \cite{PHOBOS:2010eyu} (top left panel), rapidity distributions of net protons compared to NA49 data at 17.3 GeV \cite{NA49:1998gaz} (bottom left panel) and transverse momentum spectra of positive pions, kaons, protons and anti-protons compared to STAR data \cite{STAR:2017sal} (right panels).}\label{fig1}
\vspace*{-15pt}
\end{figure*}
Due to space limitations, only a selection of basic observables pertaining to collision energy range $\sqrts=7.7\dots 19.6$~GeV is presented here. For the basic observables in a broader collision energy range the Reader is again referred to \cite{Cimerman:2023hjw}. Figure~\ref{fig1} shows pseudorapidity distributions of all charged hadrons, rapidity distributions of net protons and transverse momentum distributions of pions, kaons, protons and protons in Au-Au collisions at $\sqrts=19.6$~GeV. Due to the lack of experimental data from STAR, the rapidity distribution of net protons is compared to an older experimental data from NA49 experiment at SPS, with somewhat different collision system (Pb-Pb) at collision energy $\sqrts=17.3$~GeV per nucleon pair, which we consider being close enough to Au-Au collisions at $\sqrts=19.6$~GeV.

\begin{figure*}[h]
 \centering
 \includegraphics[width=0.75\textwidth]{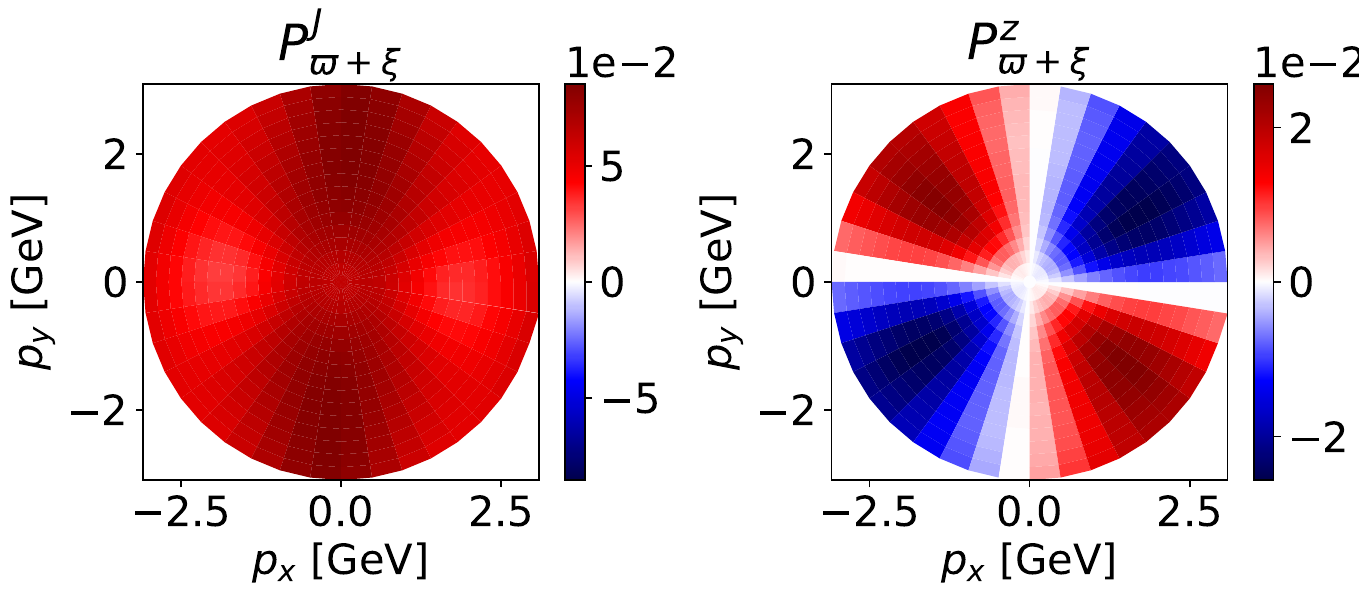}
\caption{Transverse momentum dependent polarization of $\Lambda$ hyperons from MUFFIN, computed at mid-rapidity for 20-50\% central Au-Au collisions at $\sqrts=7.7$~GeV. The panels show the polarization component along the total angular momentum (left), and along the beam direction (right).}\label{fig2}
\vspace*{-15pt}
\end{figure*}

\begin{figure}
\centering
\sidecaption
\includegraphics[width=7.5cm,clip]{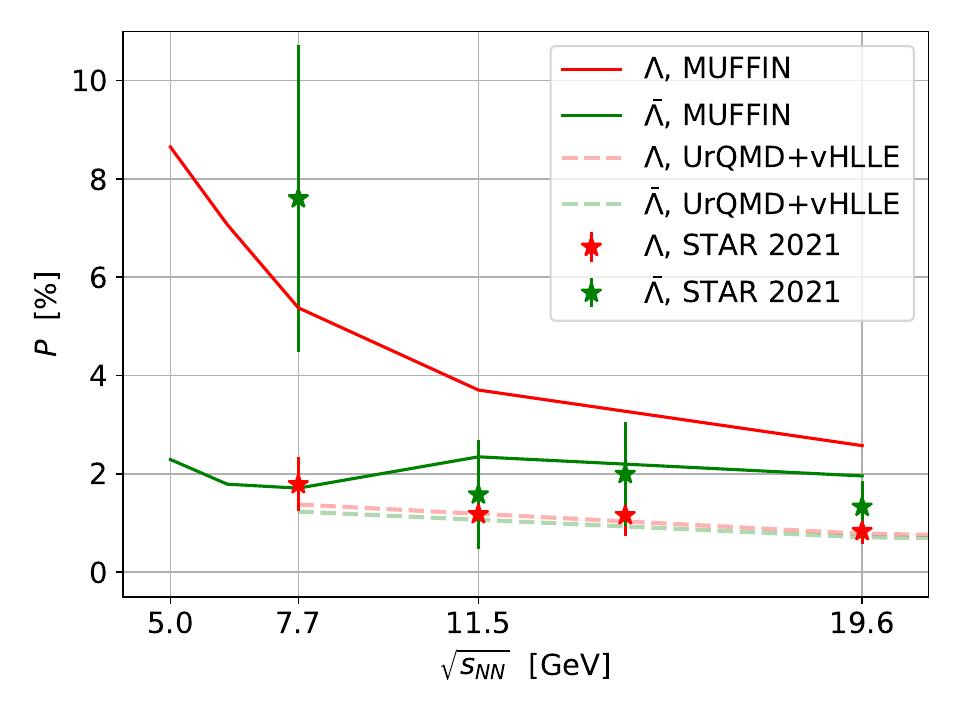}
\caption{Momentum-averaged (global) polarization of $\Lambda$ and anti-$\Lambda$ hyperons at mid-rapidity in 20-50\% central Au-Au collisions computed with MUFFIN (solid lines), compared to existing results from UrQMD+vHLLE (dashed lines) and experimental data from STAR \cite{STAR:2017ckg} corrected for updated value of $\alpha_H$ decay parameter.}
\label{fig3}
\vspace*{-10pt}
\end{figure}

Overall, MUFFIN reproduces the shapes of both the $dN_{\rm ch}/d\eta$ and net proton distributions while slightly missing the magnitudes of both. The shape of net proton distribution from NA49 experiment is reproduced well, which indicates a correct amount of baryon stopping in the present version of the friction terms. The transverse momentum distributions are reproduced generally well, with a notable exception of anti-proton ones, where the spectra are flatter than the data.

Next, we turn to hyperon polarization observables. Mean polarization of $\Lambda$ hyperons at mid-rapidity is evaluated at the hypersurface of particlization according to the formula from \cite{Becattini:2013fla}, including shear-spin coupling introduced in \cite{Becattini:2021suc}. No re-adjustment of the model parameters has been done for those observables, which can be therefore considered as a prediction from the model.

Figure~\ref{fig2} shows components of the polarization vector in the direction of the total angular momentum (on the left) and in the direction of the beam (on the right). MUFFIN simulation for this figure corresponds to 20-50\% central Au-Au collisions at $\sqrts=7.7$~GeV.

Figure~\ref{fig3} shows momentum-averaged (sometimes referred to as ``global'') polarization of $\Lambda$ and anti-$\Lambda$ hyperons, computed for 20-50\% central Au-Au collisions at different collision energies. The results from MUFFIN are shown with solid lines, whereas the results published in \cite{Karpenko:2016jyx} using conventional ``sandwich-type'' hybrid model UrQMD+vHLLE \cite{Karpenko:2015xea}. MUFFIN simulations hit the ballpark of the experimental data points and result in a strong $\Lambda$ - anti-$\Lambda$ splitting, with stronger polarization of $\Lambda$ hyperons. This seems to be in apparent contradiction to the original results from the STAR experiment \cite{STAR:2017ckg}. The updated polarization measurements presented by STAR at the SQM2024 conference, however, show a splitting which is much more consistent with zero within the error bars.

\textit{Acknowledgements:} IK and BT acknowledge support by the Czech Science Foundation under project No.~22-25026S. PH was supported by the program Excellence Initiative--Research University of the University of Wroc\l{}aw of the Ministry of Education and Science. BT acknowledges support by VEGA 1/0521/22. Computational resources were provided by the e-INFRA CZ project (ID:90254), supported by the Ministry of Education, Youth and Sports of the Czech Republic.


\vspace*{-5pt}
\begin{thebibliography}{99}
\vspace*{-5pt}

\bibitem{Cimerman:2023hjw}
J.~Cimerman, I.~Karpenko, B.~Tomasik and P.~Huovinen,
%``Next-generation multifluid hydrodynamic model for nuclear collisions at sNN from a few GeV to a hundred GeV,''
Phys. Rev. C \textbf{107}, no.4, 044902 (2023)
%doi:10.1103/PhysRevC.107.044902
[arXiv:2301.11894 [nucl-th]].

\bibitem{PHOBOS:2010eyu}
B.~Alver \textit{et al.} [PHOBOS],
%``Phobos results on charged particle multiplicity and pseudorapidity distributions in Au+Au, Cu+Cu, d+Au, and p+p collisions at ultra-relativistic energies,''
Phys. Rev. C \textbf{83}, 024913 (2011)
%doi:10.1103/PhysRevC.83.024913
[arXiv:1011.1940].

\bibitem{NA49:1998gaz}
H.~Appelshauser \textit{et al.} [NA49],
%``Baryon stopping and charged particle distributions in central Pb + Pb collisions at 158-GeV per nucleon,''
Phys. Rev. Lett. \textbf{82}, 2471-2475 (1999)
doi:10.1103/PhysRevLett.82.2471
[arXiv:nucl-ex/9810014 [nucl-ex]].

\bibitem{STAR:2017sal}
L.~Adamczyk \textit{et al.} [STAR],
%``Bulk Properties of the Medium Produced in Relativistic Heavy-Ion Collisions from the Beam Energy Scan Program,''
Phys. Rev. C \textbf{96}, no.4, 044904 (2017)
%doi:10.1103/PhysRevC.96.044904
[arXiv:1701.07065 [nucl-ex]].

\bibitem{Becattini:2013fla}
F.~Becattini, V.~Chandra, L.~Del Zanna and E.~Grossi,
%``Relativistic distribution function for particles with spin at local thermodynamical equilibrium,''
Annals Phys. \textbf{338}, 32-49 (2013)
%doi:10.1016/j.aop.2013.07.004
[arXiv:1303.3431 [nucl-th]].

\bibitem{Becattini:2021suc}
F.~Becattini, M.~Buzzegoli and A.~Palermo,
%``Spin-thermal shear coupling in a relativistic fluid,''
Phys. Lett. B \textbf{820}, 136519 (2021)
%doi:10.1016/j.physletb.2021.136519
[arXiv:2103.10917 [nucl-th]].

\bibitem{Karpenko:2016jyx}
I.~Karpenko and F.~Becattini,
%``Study of $\Lambda $ polarization in relativistic nuclear collisions at $\sqrt{s_\mathrm {NN}}=7.7$ \textendash{}200 GeV,''
Eur. Phys. J. C \textbf{77}, no.4, 213 (2017)

\bibitem{Karpenko:2015xea}
I.~A.~Karpenko, P.~Huovinen, H.~Petersen and M.~Bleicher,
%``Estimation of the shear viscosity at finite net-baryon density from $A+A$ collision data at $\sqrt{s_\mathrm{NN}} = 7.7-200$ GeV,''
Phys. Rev. C \textbf{91}, no.6, 064901 (2015)
%doi:10.1103/PhysRevC.91.064901
[arXiv:1502.01978 [nucl-th]].

\bibitem{STAR:2017ckg}
L.~Adamczyk \textit{et al.} [STAR],
%``Global $\Lambda$ hyperon polarization in nuclear collisions: evidence for the most vortical fluid,''
Nature \textbf{548}, 62-65 (2017)

\end{thebibliography}
\end{document}